\documentclass[twocolumn,showpacs,prd]{revtex4}
\usepackage{amsmath}
\usepackage{mathrsfs,bm,multirow}
\usepackage{longtable,lscape}
\usepackage{txfonts}
\usepackage{amssymb}
\usepackage{indentfirst}
\usepackage{graphicx,,booktabs}
\usepackage{color}
\usepackage{amssymb}
\begin{document}

\title{Tetraquark States with Open Flavors}
\author{Liang Tang$^{1,3}$}
\email{tangl@hebtu.edu.cn}
\author{Cong-Feng Qiao$^{2,3}$\footnote{Corresponding author}}
\email{qiaocf@ucas.ac.cn}
\affiliation{$^1$Department of Physics, Hebei Normal University, Shijiazhuang 050024, China\\
$^2$School of Physics, University of Chinese Academy of Sciences - YuQuan Road 19A, Beijing 100049, China\\
$^3$CAS Center for Excellence in Particle Physics, Beijing 100049, China}

\begin{abstract}
In this work, we estimate the masses of tetraquark states with four different flavors by virtue of QCD sum rules, in both $b$ and $c$ sectors. We construct four $[8_c]_{\bar{b} s} \otimes [8_c]_{\bar{d} u}$ tetraquark currents with $J^P = 0^+$, and then perform analytic calculation up to dimension eight in the Operator Product Expansion (OPE). We keep terms which are linear in the strange quark mass $m_s$, and in the end find two possible tetraquark states with masses $(5.57 \pm 0.15)$ and $(5.58 \pm 0.15)$ GeV. We find that their charmed-partner masses lie in $(2.54 \pm 0.13)$ and $(2.55 \pm 0.13)$ GeV, respectively and are hence accessible in experiments like BESIII and Belle.
\end{abstract}
 \maketitle
\newpage

\section{Introduction}

Recently, the D{\O} Collaboration has reported the first observation of a narrow structure, call $X(5568)$, in the decay chain $X(5568) \to B_s^0 \pi^\pm$, $B_s^0 \to J/\psi \phi$, $J/\psi \to \mu^+ \mu^-$, $\phi \to K^+ K^-$ based on the $p \bar{p}$ collision data at $\sqrt{s}= 1.96$ TeV collected at the Fermilab Tevatron collider \cite{D0:2016mwd}. Its mass and width were respectively measured to be $M_X = 5567.8 \pm 2.9 ^{+0.9}_{-1.9}$ MeV and $\Gamma_X = 21.9 \pm 6.4^{+5.0}_{-2.5}$ MeV, and the favored quantum number is $J^P = 0^+$. The statistical significance including the look-elsewhere effect and systematic errors is about 5.1 $\sigma$. The decay final state $B_s^0 \pi^+$ indicate that the component of X(5568) has to be $s u\bar{b} \bar{d}$, and this is the first observation of a hadronic state with four different flavors. However, more recently a preliminary analysis
on data collected $\sqrt{s}=7$ TeV and 8 TeV by the LHCb Collaboration does not confirm the existence of state X(5568) \cite{LHCb}. The contradictory information from D{\O} and LHCb Collaborations on X(5568) is very interesting and urges more investigations of related topics. 

The observation of X(5568) has immediately inspired extensive discussions on the possibility of its internal structure. Very recently, authors investigated the $X(5568)$ as a scalar tetraquark state using the approach of QCD sum rules \cite{Agaev:2016mjb, Wang:2016mee, Zanetti:2016wjn, Dias:2016dme, Agaev:2016ijz}. Meanwhile, in Ref.\cite{ Chen:2016mqt}, authors constructed a series of tetraquark currents to calculate the corresponding mass in the framework of QCD sum rules, and their results support the X(5568) as a tetraquark state with quantum numbers $J^P=0^+$ or $1^+$. Wang and Zhu \cite{Wang:2016tsi} employed the effective Hamiltonian approach to calculate the mass of the tetraquark state. The molecular picture of $X_b$ was carried out in Ref.\cite{Agaev:2016urs} using the QCD sum rules, where the $X_b(5568)$ was taken as the $B\bar{K}$ bound states. Assumed the X(5568) as the S-wave $B \bar{K}$ molecular state, Xiao and Chen discussed the decay of $X(5568) \to B_s^0 \pi^+$ in Ref. \cite{Xiao:2016mho}. Besides, a number of works that analyze this exotic X(5568) were accomplished based on other theoretical methods \cite{Liu:2016xly, Liu:2016ogz, He:2016yhd,Stancu:2016sfd}.

It should be noted that according to the Quantum Chromodynamics (QCD), there exists a tetraquark configuration, which is composed of two color-octet parts. Since there exits QCD interaction, it is different from the molecular state with two color-singlet mesons. That is to say, it could decay to two mesons by exchanging one or more gluons. Therefore, the study of the color-octet tetraquark state is very important for possible new hadron states. In this work, we will construct four color-octet tetraquark currents , and calculate their masses by virtue of QCD sum rules.

After the introduction, we present the primary formulas of the QCD sum rules in Sec.II. The numerical analyses and related figures are shown in Sec.III. Finally, we give a short summary of the tetraquark states with open flavors in Sec.IV.

\section{Formalism}

In this work, we study the color-octet tetraquark state with $J^P = 0^+$ via the approach of the QCD sum rules~\cite{Shifman, Reinders:1984sr, Narison:1989aq, P.Col}. The starting point of the QCD sum rules is the two-point correlation function constructed from two hadronic currents. For a scalar state that considered in this work, the two-point correlation function has the following form:
\begin{eqnarray}
\Pi(q^2) = i \int d^4 x e^{i q \cdot x} \langle 0 | T \big{\{} j(x) j^\dagger(0) \big{\}} |0 \rangle \; .
\end{eqnarray}
Here $j(x)$ and $j(0)$ are the above-mentioned hadronic currents with $J^P = 0^+$ , and they are constructed as follows:
\begin{eqnarray}
j_A(x) \!\!\! &=& \!\!\! \bigg[i \, \bar{b}^j(x) \gamma_5 (t^a)_{jk} s^k(x) \bigg] \bigg[i \, \bar{d}^m(x) \gamma_5 (t^a)_{mn} u^n(x) \bigg]\; , \label{current-1} \\
j_B(x) \!\!\! &=& \!\!\! \bigg[\bar{b}^j(x) (t^a)_{jk} s^k(x) \bigg] \bigg[ \bar{d}^m(x) (t^a)_{mn} u^n(x) \bigg]\; , \label{current-2} \\
j_C(x) \!\!\! &=& \!\!\! \bigg[ \bar{b}^j(x) \gamma^\mu (t^a)_{jk} s^k(x) \bigg] \bigg[ \bar{d}^m(x) \gamma_\mu  (t^a)_{mn} u^n(x) \bigg]\; , \label{current-3} \\
j_D(x) \!\!\! &=& \!\!\! \bigg[ \bar{b}^j(x) \gamma^\mu \gamma_5 (t^a)_{jk} s^k(x) \bigg] \bigg[ \bar{d}^m(x) \gamma_\mu \gamma_5 (t^a)_{mn} u^n(x) \bigg] , \label{current-4}
\end{eqnarray}
where $j$, $k$, $m$, and $n$ are color indices, $t^a$ is the generator of the group $SU_c(3)$. Here, the subscripts $A$, $B$, $C$, and $D$ represent the currents composed of two $0^-$ color-octet constituents, two $0^+$, two $1^-$, and two $1^+$, respectively. We will take into account all these currents in the following calculation.

The principle of quark-hadron duality is the basic assumption to employ the approach of the QCD sum rules, as shown in Ref.\cite{P.Col}. Hence, on the one hand, the correlation function $\Pi(q^2)$ can be calculated at the quark-gluon level, where the Operator Product Expansion (OPE) is employed; on the other hand, it can be expressed at the hadron level, in which the coupling constant and mass of the hadron are introduced.

In order to calculate the spectral density of the OPE side, the ``full" propagators $S^q_{i j}(x)$ and $S^Q_{i j}(p)$ of a light quark ($q=u$, $d$ or $s$) and a heavy quark ($Q=c$ or $b$) are used:
\begin{eqnarray}
S^q_{j k}(x) \! \! & = & \! \! \frac{i \delta_{j k} x\!\!\!\slash}{2 \pi^2
x^4} - \frac{\delta_{jk} m_q}{4 \pi^2 x^2} - \frac{i g_s t^a_{j k}
G^a_{\alpha \beta}}{32 \pi^2 x^2}(\sigma^{\alpha \beta} x\!\!\!\slash
+ x\!\!\!\slash \sigma^{\alpha \beta})\nonumber \\ &-& \frac{\delta_{jk}
}{12} \langle\bar{q} q \rangle + \frac{i\delta_{j k}
x\!\!\!\slash}{48} m_q \langle \bar{q}q \rangle - \frac{\delta_{j k} x^2}{192} \langle g_s \bar{q} \sigma \cdot G q \rangle \nonumber \\ &+& \frac{i \delta_{jk} x^2 x\!\!\!\slash}{1152} m_q \langle g_s \bar{q} \sigma \cdot G q \rangle - \frac{t^a_{j k} \sigma_{\alpha^\prime \beta^\prime}}{192}
\langle
g_s \bar{q} \sigma \cdot G^\prime q \rangle \nonumber \\
&+& \frac{i t^a_{jk}}{768} (\sigma_{\alpha^\prime \beta^\prime} x \!\!\!\slash + x\!\!\!\slash \sigma_{\alpha^\prime \beta^\prime}) m_q \langle
g_s \bar{q} \sigma \cdot G^\prime q \rangle \;,
\end{eqnarray}
\begin{eqnarray}
S^Q_{j k}(p) \! \! & = & \! \! \frac{i \delta_{j k}(p\!\!\!\slash + m_Q)}{p^2 - m_Q^2} - \frac{i}{4} \frac{g_s t^a_{j k} G^a_{\alpha \beta}}{(p^2 - m_Q^2)^2} [\sigma^{\alpha \beta}
(p\!\!\!\slash + m_Q) \nonumber \\
&+& (p\!\!\!\slash + m_Q) \sigma^{\alpha \beta}] + \frac{i\delta_{jk}m_Q  \langle g_s^2 G^2\rangle}{12(p^2 - m_Q^2)^3}\bigg[ 1 + \frac{m_Q (p\!\!\!\slash + m_Q)}{p^2 - m_Q^2} \bigg] \nonumber \\ &+& \frac{i \delta_{j k}}{48} \bigg\{ \frac{(p\!\!\!\slash +
m_Q) [p\!\!\!\slash (p^2 - 3 m_Q^2) + 2 m_Q (2 p^2 - m_Q^2)] }{(p^2 - m_Q^2)^6} \nonumber \\
&\times& (p\!\!\!\slash + m_Q)\bigg\} \langle g_s^3 G^3 \rangle \; ,
\end{eqnarray}
where the vacuum condensates are clearly displayed, and the Lorentz indices $\alpha^\prime$ and $\beta^\prime$ correspond to the indices of an input gluon field $G^\prime$ from another propagator\cite{Albuquerque:2013ija}.

Accordingly, based on the dispersion relation, the correlation function $\Pi(q^2)$ at the quark-gluon level can be obtained:
\begin{eqnarray}
\Pi^{OPE}_i (q^2) &=& \int_{(m_b + m_s)^2}^{\infty} d s
\frac{\rho^{OPE}_i(s)}{s - q^2} + \Pi_i^{\langle G^3 \rangle }(q^2) \nonumber \\ &+&  \Pi_i^{\langle \bar{q} q \rangle \langle \bar{q} G q \rangle}(q^2) +  \Pi_i^{\langle G^2 \rangle^2}(q^2)\; ,
\end{eqnarray}
where $\rho^{OPE}_i(s) = \text{Im} [\Pi_i^{OPE}(s)] / \pi$ and
\begin{eqnarray}
\rho^{OPE}_i(s) & = & \rho^{pert}_i(s) + \rho^{\langle \bar{s} s
\rangle}_i(s) + \rho^{\langle G^2 \rangle}_i(s) + \rho^{\langle \bar{s} G s \rangle}_i(s) + \rho^{\langle \bar{q} q \rangle^2}_i(s) \nonumber \\ &+& \rho^{\langle G^3 \rangle}_i(s) + \rho^{\langle \bar{q} q \rangle \langle \bar{q} G q \rangle}_i(s) + \rho^{\langle G^2 \rangle^2}_i (s)\; ,
\end{eqnarray}
in which the subscript $i$ runs from A to D, and $\Pi_i^{\langle G^3 \rangle }(q^2)$, $\Pi_i^{\langle \bar{q} q \rangle \langle \bar{q} G q \rangle}(q^2)$ and $\Pi_i^{\langle G^2 \rangle}(q^2)$ represent those contributions of the correlation function that do not have imaginary parts but have nontrivial values after the Borel transform.

Applying the Borel transform to the quark-gluon side, we have,
\begin{eqnarray}
\Pi_{i}^{OPE}(M_B^2) &=& \int_{(m_b + m_s)^2}^{\infty} d s
\rho^{OPE}(s)e^{- s / M_B^2} + \Pi_i^{\langle G^3 \rangle }(M_B^2) \nonumber \\ &+&  \Pi_i^{\langle \bar{q} q \rangle \langle \bar{q} G q \rangle}
(M_B^2) + \Pi_i^{\langle G^2 \rangle}(M_B^2) \; . \label{quark-gluon}
\end{eqnarray}
In order to take into account the effects induced by the mass of the strange quark, we keep terms which are linear in the strange quark mass $m_s$ in the following calculations. For all the tetraquark states considered in this article, we put the detailed formulas of spectral densities in Eq.(\ref{quark-gluon}) into the Appendix.

On the hadron side, after isolating the ground state contribution of the tetraquark state, we obtain the correlation function $\Pi(q^2)$ which is expressed as a dispersion integral over a physical regime,
\begin{eqnarray}
\Pi_i(q^2) & = & \frac{(\lambda_{X}^{i})_2}{(M_{X}^i)^2 - q^2} + \frac{1}{\pi} \int_{s_0}^\infty d s \frac{\rho^{i}_X(s)}{s - q^2} \; , \label{hadron}
\end{eqnarray}
where $M_{X}^i$ is the mass of the tetraquark state with $J^P = 0^+$, and $\rho^{i}_X(s)$ is the spectral density that contains the contributions from the higher excited states and the continuum states, $s_0$ is the threshold of the higher excited states and continuum states. The coupling constant $\lambda_{X}$ is defined by $\langle 0 | j_X^i | X \rangle = \lambda_{X}^i$, where $X$ is the lowest lying tetraquark state.

Performing the Borel transform on the hadron side (Eq.(\ref{hadron})) and then matching it to Eq.(\ref{quark-gluon}), we can obtain the mass of the scalar tetraqark state with open flavors:
\begin{eqnarray}
M_{X}^i(s_0, M_B^2) = \sqrt{- \frac{R_1^i(s_0, M_B^2)}{R_0^i(s_0, M_B^2)}} \; , \label{mass-Eq}
\end{eqnarray}
where $X$ denotes the tetraquark state and
\begin{eqnarray}
R_0^i(s_0, M_B^2) & = & \int_{(m_b + m_s)^2}^{s_0} d s \; \rho^{OPE}(s) e^{-
s / M_B^2} + \Pi_i^{\langle G^3 \rangle }(M_B^2) \nonumber \\
& &  + \Pi_i^{\langle \bar{q} q \rangle \langle  \bar{q} G q \rangle}(M_B^2) + \Pi_i^{\langle G^2\rangle^2}(M_B^2)\; , \\ R_1^i(s_0, M_B^2) & = &
\frac{\partial}{\partial{M_B^{-2}}}{R_0(s_0, M_B^2)} \; .
\end{eqnarray}

\section{Numerical Results}

The expressions of the QCD sum rules contain various input parameters, such as the the condensates and the
quark masses. As shown in Refs.\cite{Shifman, Reinders:1984sr, P.Col, Chen:2016mqt, Narison:1989aq, Zanetti:2016wjn, Agashe:2014kda}, we take these values as:$m_u = m_d =0$,
$m_s(2 \text{GeV}) = (95 \pm 5) \; \text{MeV}$, $m_c (m_c) = \overline{m}_c= (1.275 \pm 0.025)
\; \text{GeV}$, $m_b (m_b) = \overline{m}_b= (4.18 \pm 0.03)
\; \text{GeV}$, $\langle \bar{q} q \rangle = - (0.24 \pm 0.01)^3
\; \text{GeV}^3$, $\langle \bar{s} s \rangle = (0.8 \pm 0.1)
\langle \bar{q} q \rangle$, $\langle g_s^2 G^2 \rangle = 0.88
\; \text{GeV}^4$, $\langle \bar{s} g_s \sigma \cdot G s
\rangle = m_0^2 \langle \bar{s} s \rangle$, $\langle g_s^3 G^3
\rangle = 0.045 \; \text{GeV}^6$, and $m_0^2 = 0.8 \; \text{GeV}^2$.
Here, $\overline{m}_c$ and $\overline{m_b}$ are the running masses of the heavy quarks in the $\overline{MS}$ scheme.

Moreover, there exist two additional parameters $M_B^2$ and $s_0$ introduced by the QCD sum rules, which should be fixed in accordance with the standard procedures. In Refs.\cite{Shifman, Reinders:1984sr, P.Col}, there are two criteria to constrain the parameter $M_B^2$ and the threshold $s_0$. The first criteria is the convergence of the OPE. That is, we need to compare the relative contribution of each term to the total contributions of the OPE side, and choose a reliable region of $M_B^2$ to retain their convergence. Second, the pole contribution (PC) defined as the pole contribution (corresponding to the contribution of the ground state) divided by the total contribution (pole plus continuum), should be larger than 50\%~\cite{P.Col, Matheus:2006xi}. Thus, we can safely eliminate the contributions of the higher excited and continuum states.

\begin{figure}[t]
\includegraphics[width=5.8cm]{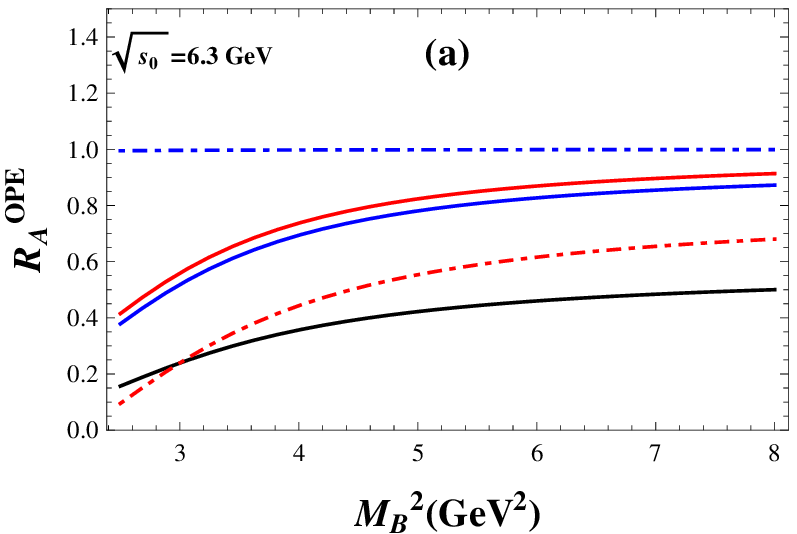}
\includegraphics[width=5.8cm]{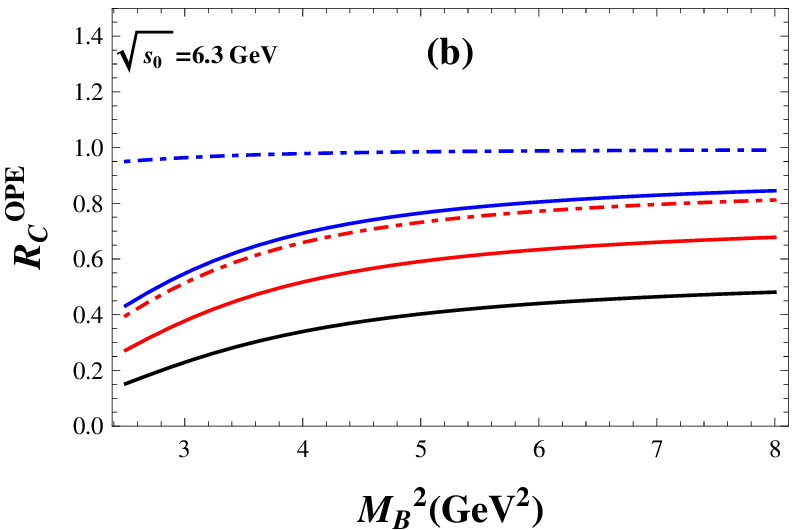}
\caption{(Color online) (\textbf{a}) The OPE convergence $R^{OPE}_A$ as a function of the Borel parameter $M_B^2$ in the region $2.5 \leq M_B^2 \leq 8.0 \; \text{GeV}^{2}$ for the tetraquark state of case $A$, where $\sqrt{s_0} = 6.3 \; \text{GeV}$.  (\textbf{b}) The OPE convergence $R^{OPE}_C$ as a function of the Borel parameter $M_B^2$ in the region $2.5 \leq M_B^2 \leq 8.0 \; \text{GeV}^{2}$ for the tetraquark state of case $C$, where $\sqrt{s_0} = 6.3 \; \text{GeV}$. The black line represents the fraction of perturbative contribution, and each subsequent line stands for the addition of one extra condensate, {\it i.e.}, $+ \langle \bar{s} s \rangle$ (red line), $+ \langle g_s^2 G^2 \rangle$ (blue line), $+ \langle g_s \bar{s} \sigma \cdot G s \rangle$  (red dotted line), $+ \langle \bar{q} q \rangle^2$ (blue dotted line). Since the curves that add the condensate terms of $\langle g_s^3 G^3 \rangle$, $\langle \bar{q} q \rangle \langle g_s \bar{q} \sigma \cdot G q \rangle$ and $\langle g_s^2 G^2 \rangle^2$ one by one are just straight lines, respectively, therefore we do not show them here.} \label{fig1}
\end{figure}

Meanwhile, in order to find a proper value of $\sqrt{s_0}$, we perform a similar analysis as in Refs.~\cite{Finazzo:2011he, Qiao:2013dda}. Because the continuum threshold $s_0$ is connected to the mass of the ground state by the relation $\sqrt{s_0} \sim (M_X + \delta) \, \text{GeV}$, in which $\delta$ lies in the range of $0.4 \sim 0.8$ GeV, various $\sqrt{s_0}$ satisfying this constraint should be taken into account in the numerical analyses. Among these values, we need then to pick out the proper one that has an optimal window for Borel parameter $M_B^2$. That is to say, in this optimal window, the tetraquark mass $M_X$ is independent of the Borel parameter $M_B^2$ as much as possible. Finally, the value of $\sqrt{s_0}$ corresponding to the optimal mass curve is the central value of $\sqrt{s_0}$. In practice, it is normally acceptable to vary the $\sqrt{s_0}$ by $0.2$ GeV~\cite{Qiao:2013raa, Qiao:2013xca} in the QCD sum rules calculation, which determines the upper and lower bounds of $\sqrt{s_0}$. Hence, these bounds give rise to the uncertainties of $\sqrt{s_0}$.

\begin{figure}[t]
\begin{center}
\includegraphics[width=5.8cm]{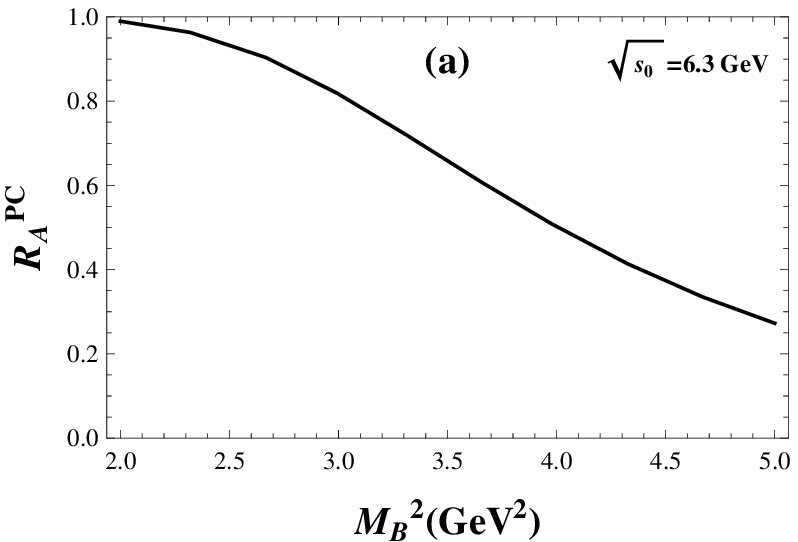}
\includegraphics[width=5.8cm]{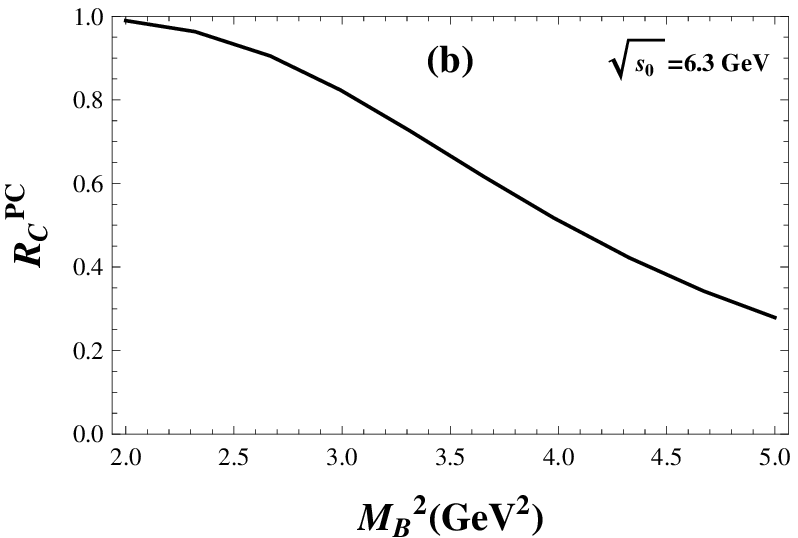}
\caption{(\textbf{a}) The pole contribution $R^{PC}_A$ for the tetraquark state as a function of the Borel parameter $M_B^2$ in case $A$ with $\sqrt{s_0} = 6.3 \; \text{GeV}$. (\textbf{b}) The pole contribution $R^{PC}_C$ for the tetraquark state as a function of the Borel parameter $M_B^2$ in case $C$ with $\sqrt{s_0} = 6.3 \; \text{GeV}$.} \label{fig2}
\end{center}
\end{figure}

\begin{figure}[t]
\begin{center}
\includegraphics[width=5.8cm]{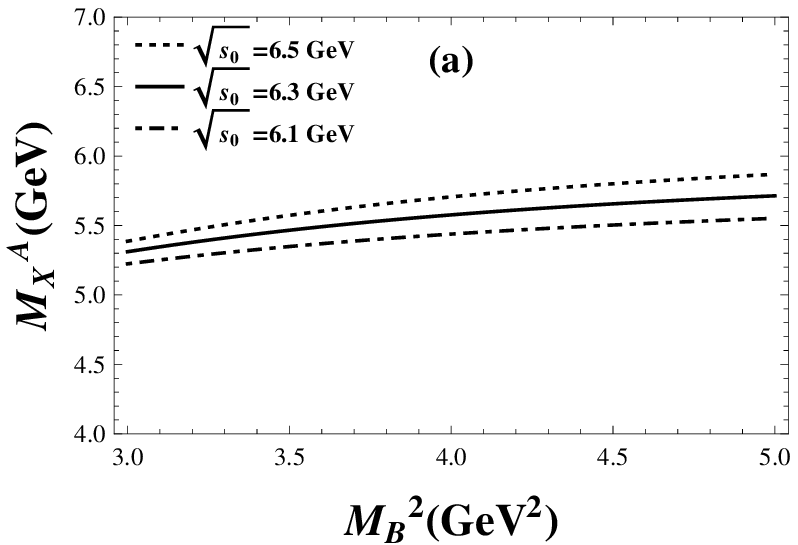}
\includegraphics[width=5.8cm]{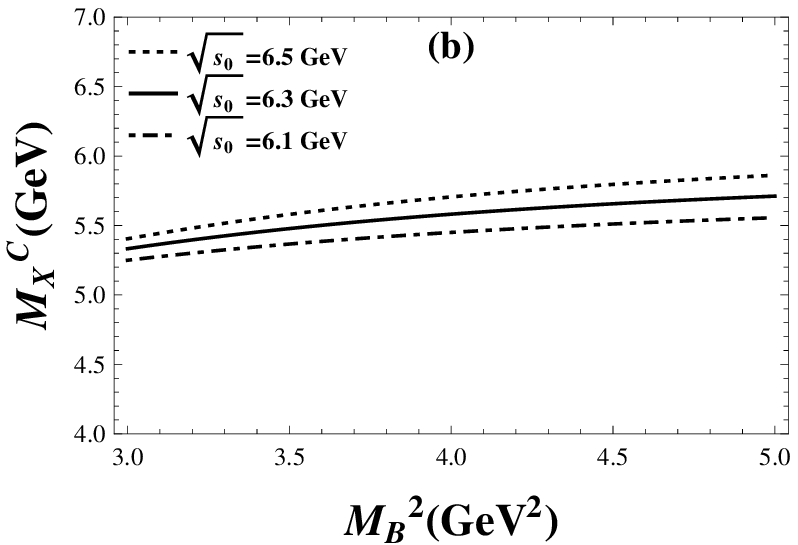}
\caption{(\textbf{a}) The mass of the tetraquark state in case $A$ as a function of the Borel parameter $M_B^2$, for different values of $\sqrt{s_0}$. (\textbf{b}) The mass of the tetraquark state in case $C$ as a function of the Borel parameter $M_B^2$, for different values of $\sqrt{s_0}$. } \label{fig3}
\end{center}
\end{figure}

We illustrate the OPE convergences in Figs.(\ref{fig1}-a, \ref{fig1}-b) respectively for case $A$ and $C$. Performing the first criterion, we find the lower limit constraint of $M_B^2$ is $M_B^2 \gtrsim 3.0 \; \text{GeV}^{2}$  with $\sqrt{s_0} = 6.3 \, \text{GeV}$ for both case $A$ and $C$. The curve of the pole contribution $R^{PC}$ are drawn in Figs.(\ref{fig2}-a,\ref{fig2}-b), which indicate the upper limit constraint of $M_B^2$ is $M_B^2 \lesssim 4.0 \; \text{GeV}^{2}$ with $\sqrt{s_0} = 6.3 \, \text{GeV}$ for both case $A$ and $C$. It should be noted that the limit constraints of $M_B^2$ also depend on the threshold parameter $\sqrt{s_0}$. That is, there are different limit constraints of $M_B^2$ for different $\sqrt{s_0}$. For determining the value of $\sqrt{s_0}$, we carry out an analysis similar to Ref.~\cite{Matheus:2006xi}. The masses $M_X^A$ and $M_X^C$ as a function of the Borel parameter $M_B^2$ for different values $\sqrt{s_0}$ are drawn in Figs.(\ref{fig3}-a,\ref{fig3}-b).

However, we find there doesn't exist a reasonable region of the parameter $M_B^2$ for both case $B$ and $D$. Therefore, we can conclude that these two cases do not correspond to any tetraquark states.

Eventually, the masses of the tetraquark states with currents $A$ and $C$ are determined to be
\begin{eqnarray}
M_X^A &=& (5.57 \pm 0.15) \, \text{GeV} \; , \\
M_X^C &=& (5.58 \pm 0.15) \, \text{GeV} \; ,
\end{eqnarray}
where the central value of the mass $M_X$ corresponds to the result with the optimal stability of $M_B^2$, and the errors stem from the uncertainties of the condensates, the quark mass, the threshold parameter $\sqrt{s_0}$ and the Borel parameter $M_B^2$.

Moreover, we predict their charmed partners with masses of $(2.54 \pm 0.13)$ and $(2.55 \pm 0.13)$ GeV, respectively.

\section{Summary}
In this work, we estimate the masses of tetraquark states with four different flavors by virtue of QCD sum rules, in both $b$ and $c$ sectors. We construct four $[8_c]_{\bar{b} s} \otimes [8_c]_{\bar{d} u}$ tetraquark currents with $J^P = 0^+$, and then perform analytic calculation up to dimension eight in the OPE. We keep terms which are linear in the strange quark mass $m_s$.

The numerical results are respectively $(5.57 \pm 0.15) \, \text{GeV}$ and $(5.58 \pm 0.15) \, \text{GeV}$ for case $A$ and $C$. However, due to the lack of reasonable windows of the Borel parameter $M_B^2$,  case $B$ and $D$ do not correspond to any hadron states. Our results imply that two S-wave octet parts can form a resonance, whereas two P-wave octet parts can not form a resonance. Therefore, we can speculate that two $[8_c]_{\bar{b} s} \otimes [8_c]_{\bar{d} u}$ tetraquark states with $J^P = 1^+$ should exist and have a degenerate mass. We will present detailed analyses of tetraquark states with $J^P = 1^+$ in our next work.

In conclusion, we find in b-quark sector two possible open-flavor tetraquark states with masses $(5.57 \pm 0.15)$ and $(5.58 \pm 0.15)$ GeV may exist, while their charmed-partners lie in $(2.54 \pm 0.13)$ and $(2.55 \pm 0.13)$ GeV, respectively and are hence accessible in experiments like BESIII and Belle. Though a preliminary analysis performed by the LHCb collaboration does not favor of the existence of $X(5568)$ \cite{LHCb}, the tetraquark state with open flavors is still an interesting target deserving more explorations.

%%%%%%%%%%%%%%%%%%%%%%%%%%%%%%%%%%%%%%%%%%%%%%%%%%%%%%%%%%%%%%%%%%%%%%
\vspace{.7cm} {\bf Acknowledgments} \vspace{.3cm}

This work was supported in part by the Ministry of Science and Technology of the People's Republic of China (2015CB856703), by Science Foundation of Hebei Normal University under Contract No. L2016B08, and by the National Natural Science Foundation of China(NSFC) under the grants 11375200 and 11547190.

%%%%%%%%%%%%%%%%%%%%%%%%%%%%%%%%%%%%%%%%%%%%%%%%%%%%%%%%%%%%%%%%%%%%%%%

\begin{widetext}
\appendix

\section{The spectral densities for cases A to D}

For case $A$ where the current is composed of two $0^-$ color-octet parts, we obtain the spectral density as follows:
\begin{eqnarray}
\rho^{pert}_A (s) &=&  \frac{1}{2^{11} \times 3^2 \pi^6} \int_0^\lambda d \alpha \frac{H_\alpha^3 (H_\alpha - 4 m_b m_s \alpha)}{\alpha (1 - \alpha)^3} \, , \\
\rho^{\langle \bar{s} s \rangle}_A (s) &=& \frac{\langle \bar{s} s \rangle}{2^7 \times 3 \pi ^4} \int_0^\lambda d \alpha \frac{H_\alpha^2 (m_s - \alpha (m_b + m_s))}{\alpha (1 - \alpha)^2} \, , \\
\rho^{\langle GG \rangle}_A (s) & = & - \frac{\langle g_s^2 GG\rangle}{2^{15} \times 3^3 \pi^6} \int_0^\lambda d \alpha \frac{1}{\alpha (1 - \alpha)^3} [9 H_\alpha^2 (\alpha^2 - 3\alpha +2)  \nonumber \\ & & - 4 H_\alpha m_b \alpha (4 m_b \alpha^2 - 3 m_s (\alpha^2 + 6\alpha -3)) + 16 m_b^3 m_s \alpha^4] \, , \\
\rho^{\langle \bar{s} G s\rangle}_A (s) &= & \frac{\langle \bar{s} g_s \sigma \cdot G s\rangle}{2^{10} \times 3^2 \pi^4} \int_0^\lambda d \alpha \frac{H_\alpha [3m_b \alpha(7\alpha -8) + m_s (5\alpha^2 -13 \alpha + 8)]}{\alpha (1 - \alpha)^2} \, ,
\end{eqnarray}
\begin{eqnarray}
\rho^{\langle G^3\rangle}_A (s) &=& \frac{\langle g_s^3 G^3\rangle}{2^{13} \times 3^3 \pi^6} \int_0^\lambda d \alpha \frac{\alpha^2 (H_\alpha + 2 m_b \alpha (m_b - 3 m_s))}{(1 - \alpha)^3} \, , \\
\rho^{\langle \bar{q} q \rangle^2}_A (s) & = & \frac{\langle \bar{q} q \rangle }{72 \pi^2} \int_0^\lambda d \alpha \frac{(m_b m_s \alpha - H_\alpha)}{\alpha} \, , \\
\rho^{\langle G^2 \rangle^2}_A (s) & = & \frac{\langle g_s^2 G^2 \rangle^2}{2^{16} \times 3 \pi^6} \lambda \, , \\
\rho^{\langle \bar{q} q\rangle \langle \bar{q} G q\rangle}_A(s) &=& 0 \, ,
\end{eqnarray}
\begin{eqnarray}
\Pi^{\langle G^3\rangle}_A (M_B^2) &=& \frac{\langle g_s^3 G^3\rangle}{2^{12}\times 3^3 \pi^6} \int_0^1 d \alpha \frac{m_b^3 m_s \alpha^3}{(1 - \alpha)^4} e^{- \frac{m_b^2}{\alpha (1 - \alpha) M_B^2}} \, , \\
\Pi^{\langle \bar{q} q \rangle \langle \bar{q} G q \rangle}_A (M_B^2) &=& - \frac{\langle \bar{q} q \rangle \langle \bar{q} g_s \sigma \cdot G q \rangle}{2^4 \times 3^2 \pi^2} \int_0^1 d \alpha \frac{1}{\alpha^2 (1 - \alpha)}e^{- \frac{m_b^2}{\alpha (1 - \alpha) M_B^2}} \, , \\
\Pi^{\langle G^2 \rangle^2}_A (M_B^2) &=& \frac{\langle g_s^2 G^2 \rangle^2}{2^{16} \times 3^3 \pi^6} \int_0^1 d \alpha \frac{m_b \alpha}{(1 - \alpha)^3 M_B^2} \bigg[ m_b^2 m_s + M_B^2 (1 -  \alpha)(m_b - 3 m_s) \bigg] e^{- \frac{m_b^2}{\alpha (1 - \alpha) M_B^2}} \, ,
\end{eqnarray}
where $M_B^2$ is the Borel parameter, $H_\alpha = m_b^2 \alpha - \alpha (1 - \alpha) s$, and $\lambda = 1 - m_b^2/s$.

For case $B$ where the current is composed of two $0^+$ color-octet parts, we obtain the spectral density as follows:
\begin{eqnarray}
\rho^{pert}_B (s) &=&  \frac{1}{2^{11} \times 3^2 \pi^6} \int_0^\lambda d \alpha \frac{H_\alpha^3 (H_\alpha + 4 m_b m_s \alpha)}{\alpha (1 - \alpha)^3} \, , \\
\rho^{\langle \bar{s} s \rangle}_B (s) &=& \frac{\langle \bar{s} s \rangle}{2^7 \times 3 \pi ^4} \int_0^\lambda d \alpha \frac{H_\alpha^2 (m_s + \alpha (m_b - m_s))}{\alpha (1 - \alpha)^2} \, , \\
\rho^{\langle GG \rangle}_B (s) & = & - \frac{\langle g_s^2 GG\rangle}{2^{15} \times 3^3 \pi^6} \int_0^\lambda d \alpha \frac{1}{\alpha (1 - \alpha)^3} [9 H_\alpha^2 (\alpha^2 - 3\alpha +2) \nonumber \\ & & - 4 H_\alpha m_b \alpha (4 m_b \alpha^2 + 3 m_s (\alpha^2 + 6\alpha -3)) - 16 m_b^3 m_s \alpha^4] \, ,
\end{eqnarray}
\begin{eqnarray}
\rho^{\langle \bar{s}  G s\rangle}_B (s) &= & \frac{\langle \bar{s} g_s \sigma \cdot G s\rangle}{2^{10} \times 3^2 \pi^4} \int_0^\lambda d \alpha \frac{H_\alpha [3 m_b (8 - 7\alpha)\alpha + m_s (5\alpha^2 -13 \alpha + 8 )]}{\alpha (1 - \alpha)^2} \, , \\
\rho^{\langle G^3\rangle}_B (s) &=& \frac{\langle g_s^3 G^3\rangle}{2^{13} \times 3^3 \pi^6} \int_0^\lambda d \alpha \frac{\alpha^2 (H_\alpha + 2 m_b \alpha (m_b + 3 m_s))}{(1 - \alpha)^3} \, , \\
\rho^{\langle \bar{q} q \rangle^2}_B (s) & = & \frac{\langle \bar{q} q \rangle }{72 \pi^2} \int_0^\lambda d \alpha \frac{(m_b m_s \alpha + H_\alpha)}{\alpha} \, , \\
\rho^{\langle G^2 \rangle^2}_B (s) & = & \frac{\langle g_s^2 G^2 \rangle^2}{2^{16} \times 3 \pi^6} \lambda \, , \\
\rho^{\langle \bar{q} q\rangle \langle \bar{q} G q\rangle}_B(s) &=& 0 \, ,
\end{eqnarray}
\begin{eqnarray}
\Pi^{\langle G^3\rangle}_B (M_B^2) &=& - \frac{\langle g_s^3 G^3\rangle}{2^{12}\times 3^3 \pi^6} \int_0^1 d \alpha \frac{m_b^3 m_s \alpha^3}{(1 - \alpha)^4} e^{- \frac{m_b^2}{\alpha (1 - \alpha) M_B^2}} \, , \\
\Pi^{\langle \bar{q} q \rangle \langle \bar{q} G q \rangle}_B (M_B^2) &=& \frac{\langle \bar{q} q \rangle \langle \bar{q} g_s \sigma \cdot G q \rangle}{2^4 \times 3^2 \pi^2} \int_0^1 d \alpha \frac{1}{\alpha^2 (1 - \alpha)} e^{- \frac{m_b^2}{\alpha (1 - \alpha) M_B^2}} \, , \\
\Pi^{\langle G^2 \rangle^2}_B (M_B^2) &=& - \frac{\langle g_s^2 G^2 \rangle^2}{2^{16} \times 3^3 \pi^6} \int_0^1 d \alpha \frac{m_b \alpha}{(1 - \alpha)^3 M_B^2} \bigg[ m_b^2 m_s - M_B^2 (1 -  \alpha)(m_b + 3 m_s) \bigg] e^{- \frac{m_b^2}{\alpha (1 - \alpha) M_B^2}} \, .
\end{eqnarray}

For case $C$ where the current is composed of two $1^-$ color-octet parts, we obtain the spectral density as follows:
\begin{eqnarray}
\rho^{pert}_C (s) &=&  \frac{1}{2^{9} \times 3^2 \pi^6} \int_0^\lambda d \alpha \frac{H_\alpha^3 (H_\alpha - 2 m_b m_s \alpha)}{\alpha (1 - \alpha)^3} \, , \\
\rho^{\langle \bar{s} s \rangle}_C (s) &=& \frac{\langle \bar{s} s \rangle}{2^6 \times 3 \pi ^4} \int_0^\lambda d \alpha \frac{H_\alpha^2 ( 2 m_s(1 - \alpha) - m_b \alpha )}{\alpha (1 - \alpha)^2} \, , \\
\rho^{\langle GG \rangle}_C (s) & = & - \frac{\langle g_s^2 GG\rangle}{2^{14} \times 3^3 \pi^6} \int_0^\lambda d \alpha \frac{1}{\alpha (1 - \alpha)^3} [ -81 H_\alpha^2 (\alpha^2 - 3\alpha +2)  \nonumber \\ & & + 2 H_\alpha m_b \alpha (-16 m_b \alpha^2 + m_s (-39 \alpha^2 + 45\alpha +18)) + 16 m_b^3 m_s \alpha^4] \, , \\
\rho^{\langle \bar{s} G s\rangle}_C (s) &= & \frac{\langle \bar{s} g_s \sigma \cdot G s\rangle}{2^9 \times 3^2 \pi^4} \int_0^\lambda d \alpha \frac{H_\alpha (3m_b \alpha + 11m_s (\alpha -1))}{\alpha (1 - \alpha)} \, ,
\end{eqnarray}
\begin{eqnarray}
\rho^{\langle G^3\rangle}_C (s) &=& \frac{\langle g_s^3 G^3\rangle}{2^{11} \times 3^3 \pi^6} \int_0^\lambda d \alpha \frac{\alpha^2 (H_\alpha + m_b \alpha (2 m_b - 3 m_s))}{(1 - \alpha)^3} \, , \\
\rho^{\langle \bar{q} q \rangle^2}_C (s) & = & \frac{\langle \bar{q} q \rangle }{36 \pi^2} \int_0^\lambda d \alpha \frac{(2m_b m_s \alpha - H_\alpha)}{\alpha} \, , \\
\rho^{\langle G^2 \rangle^2}_C (s) & = & \frac{139 \langle g_s^2 G^2 \rangle^2}{2^{17} \times 3^3 \pi^6} \lambda \, , \\
\rho^{\langle \bar{q} q\rangle \langle \bar{q} G q\rangle}_C(s) &=& \frac{\langle \bar{q} q\rangle \langle \bar{q} g_s \sigma \cdot G q\rangle}{128 \pi^2} \lambda \, ,
\end{eqnarray}
\begin{eqnarray}
\Pi^{\langle G^3\rangle}_C (M_B^2) &=& \frac{\langle g_s^3 G^3\rangle}{2^{11}\times 3^3 \pi^6} \int_0^1 d \alpha \frac{m_b^3 m_s \alpha^3}{(1 - \alpha)^4} e^{- \frac{m_b^2}{\alpha (1 - \alpha) M_B^2}} \, , \\
\Pi^{\langle \bar{q} q \rangle \langle \bar{q} G q \rangle}_C (M_B^2) &=& \frac{\langle \bar{q} q \rangle \langle \bar{q} g_s \sigma \cdot G q \rangle}{2^7 \times 3^2 \pi^2} \int_0^1 d \alpha \frac{(18 m_b m_s \alpha^2 - 7)}{\alpha^2 (1 - \alpha)}e^{- \frac{m_b^2}{\alpha (1 - \alpha) M_B^2}} \, , \\
\Pi^{\langle G^2 \rangle^2}_C (M_B^2) &=& - \frac{\langle g_s^2 G^2 \rangle^2}{2^{15} \times 3^3 \pi^6} \int_0^1 d \alpha \frac{m_b \alpha}{(1 - \alpha)^3 M_B^2} \bigg[ m_b^2 m_s + M_B^2 (1 -  \alpha)(9 m_b - 3 m_s) \bigg] e^{- \frac{m_b^2}{\alpha (1 - \alpha) M_B^2}} \, .
\end{eqnarray}

For case $D$ where the current is composed of two $1^+$ color-octet parts, we obtain the spectral density as follows:
\begin{eqnarray}
\rho^{pert}_D (s) &=&  \frac{1}{2^{9} \times 3^2 \pi^6} \int_0^\lambda d \alpha \frac{H_\alpha^3 (H_\alpha + 2 m_b m_s \alpha)}{\alpha (1 - \alpha)^3} \, , \\
\rho^{\langle \bar{s} s \rangle}_D (s) &=& \frac{\langle \bar{s} s \rangle}{2^6 \times 3 \pi ^4} \int_0^\lambda d \alpha \frac{H_\alpha^2 ( 2 m_s(1 - \alpha) + m_b \alpha )}{\alpha (1 - \alpha)^2} \, , \\
\rho^{\langle GG \rangle}_D (s) & = &  \frac{\langle g_s^2 GG\rangle}{2^{14} \times 3^3 \pi^6} \int_0^\lambda d \alpha \frac{1}{\alpha (1 - \alpha)^3} [81 H_\alpha^2 (\alpha^2 - 3\alpha +2)  \nonumber \\ & & + 2 H_\alpha m_b \alpha (16 m_b \alpha^2 + m_s (-39 \alpha^2 +45 \alpha + 18)) + 16 m_b^3 m_s \alpha^4] \, , \\
\rho^{\langle \bar{s} G s\rangle}_D (s) &= & - \frac{\langle \bar{s} g_s \sigma \cdot G s\rangle}{2^9 \times 3^2 \pi^4} \int_0^\lambda d \alpha \frac{H_\alpha ( 3m_b \alpha - 11m_s (\alpha -1))}{\alpha (1 - \alpha)} \, ,
\end{eqnarray}
\begin{eqnarray}
\rho^{\langle G^3\rangle}_D (s) &=& \frac{\langle g_s^3 G^3\rangle}{2^{11} \times 3^3 \pi^6} \int_0^\lambda d \alpha \frac{\alpha^2 (H_\alpha + m_b \alpha (2 m_b + 3 m_s))}{(1 - \alpha)^3} \, , \\
\rho^{\langle \bar{q} q \rangle^2}_D (s) & = & \frac{\langle \bar{q} q \rangle }{36 \pi^2} \int_0^\lambda d \alpha \frac{(2m_b m_s \alpha + H_\alpha)}{\alpha} \, , \\
\rho^{\langle \bar{q} q\rangle \langle \bar{q} G q\rangle}_D(s) &=& - \frac{\langle \bar{q} q\rangle \langle \bar{q} g_s \sigma \cdot G q\rangle}{128 \pi^2} \lambda \, ,\\
\rho^{\langle G^2 \rangle^2}_D (s) & = & \frac{139 \langle g_s^2 G^2 \rangle^2}{2^{17} \times 3^3 \pi^6} \lambda \, ,
\end{eqnarray}
\begin{eqnarray}
\Pi^{\langle G^3\rangle}_D (M_B^2) &=& - \frac{\langle g_s^3 G^3\rangle}{2^{11}\times 3^3 \pi^6} \int_0^1 d \alpha \frac{m_b^3 m_s \alpha^3}{(1 - \alpha)^4} e^{- \frac{m_b^2}{\alpha (1 - \alpha) M_B^2}} \, , \\
\Pi^{\langle \bar{q} q \rangle \langle \bar{q} G q \rangle}_D (M_B^2) &=& \frac{\langle \bar{q} q \rangle \langle \bar{q} g_s \sigma \cdot G q \rangle}{2^7 \times 3^2 \pi^2} \int_0^1 d \alpha \frac{(18 m_b m_s \alpha^2 + 7)}{\alpha^2 (1 - \alpha)}e^{- \frac{m_b^2}{\alpha (1 - \alpha) M_B^2}} \, , \\
\Pi^{\langle G^2 \rangle^2}_D (M_B^2) &=& \frac{\langle g_s^2 G^2 \rangle^2}{2^{15} \times 3^3 \pi^6} \int_0^1 d \alpha \frac{m_b \alpha}{(1 - \alpha)^3 M_B^2} \bigg[ m_b^2 m_s - M_B^2 (1 -  \alpha)(9 m_b + 3 m_s) \bigg] e^{- \frac{m_b^2}{\alpha (1 - \alpha) M_B^2}} \, .
\end{eqnarray}
\end{widetext}

\end{document}